\documentclass{Interspeech2024}

\usepackage{amsmath,graphicx, amssymb}
\usepackage{float}
\usepackage{cite}
\usepackage{mathtools}

\usepackage{capt-of}
\usepackage{url}
\usepackage[nolist]{acronym}
\usepackage{csquotes}
\usepackage{amssymb,multirow}
\usepackage{booktabs, bm}
\usepackage{booktabs,caption,fixltx2e}
\usepackage{subfigure}
\usepackage{tikz}
\usetikzlibrary{fit,tikzmark,calc}
\usepackage{siunitx}
\usepackage{lineno}
\usepackage[]{algorithm2e}
\usepackage{hyperref}
\hypersetup{
    colorlinks=true,
    linkcolor=purple,
    filecolor=magenta,      
    urlcolor=purple,
}

\usepackage{bibspacing}
\usepackage{caption}
\usepackage{enumitem}

\usepackage{amsmath}

\newcommand\eg{\textit{e.g.}}

\newacro{MVDR}[MVDR]{Minimum Variance Distortion-less Response}
\newacro{GEV}[GEV]{Generalized Eigenvalue Decomposition}
\newacro{MCWF}[MCWF]{Multi-Channel Wiener Filter}
\newacro{SDW-MWF}[SDW-MWF]{Speech Distortion Weighted MWF}
\newacro{SISDR}[SI-SDR]{Scale Invariant Signal-to-Distortion Ratio}
\newacro{STFT}[STFT]{Short-Time Fourier Transform}
\newacro{DNN}[DNN]{Deep Neural Network}
\newacro{SNR}[SNR]{Signal-to-Noise-Ratio}
\newacro{FLOPs}[FLOPs]{Floating Point Operations}
\newacro{SIR}[SIR]{Signal-to-Interference Ratio}
\newacro{ASR}[ASR]{Automatic Speech Recognition}
\newacro{IBM}[IBM]{Ideal Binary Mask}
\newacro{IRM}[IRM]{Ideal Ratio Mask}
\newacro{WLM}[WLM]{Wiener-Like Mask}
\newacro{SDR}[SDR]{Signal-to-Distortion Ratio}
\newacro{STOI}[STOI]{Short-time Objective Intelligibility}
\newacro{MSE}[MSE]{Mean-Squared Error}
\newacro{WER}[WER]{Word Error Rate}
\newacro{WPE}[WPE]{Weighted Prediction Error}
\newacro{MISO}[MISO]{Multiple Input Single Output}
\newacro{DFL}[DFL]{Deep Feature Loss}
\newacro{DASR}[DASR]{distant automatic speech recognition}

\usepackage{multirow, multicol}
\usepackage[nolist]{acronym}
\usepackage{cite}
\usepackage{adjustbox}
\usepackage{listings}
\usepackage{xcolor}

\definecolor{codegreen}{rgb}{0,0.6,0}
\definecolor{codegray}{rgb}{0.5,0.5,0.5}
\definecolor{codepurple}{rgb}{0.58,0,0.82}
\definecolor{backcolour}{rgb}{0.95,0.95,0.92}

\interspeechfinaltrue
\title{The CH\-iME-8 DASR Challenge for Generalizable and Array Agnostic Distant Automatic Speech Recognition and Diarization}

\name[affiliation={1}]{Samuele}{Cornell}
\name[affiliation={2}]{Taejin}{Park}
\name[affiliation={2}]{Steve}{Huang}
\name[affiliation={3}]{Christoph}{Boeddeker} 
\name[affiliation={1}]{Xuankai}{Chang}
\name[affiliation={4}]{Matthew}{Maciejewski}
\name[affiliation={4}]{Matthew}{Wiesner}
\name[affiliation={4}]{Paola}{Garcia}
\name[affiliation={1}]{Shinji}{Watanabe}

\address{
  $^1$Carnegie Mellon University, USA
  $^2$NVIDIA, USA, 
  $^3$ Paderborn University, Germany \\
  $^4$ Johns Hopkins University, USA
  }
\email{samuele.cornell@ieee.org}


\begin{document}

\maketitle

\begin{abstract}
This paper presents the CH\-iME-8 DASR challenge which carries on from the previous edition CH\-iME-7 DASR (C7DASR) and the past CH\-iME-6 challenge. 
It focuses on joint multi-channel distant speech recognition (DASR) and diarization with one or more, possibly heterogeneous, devices. 
The main goal is to spur research towards meeting transcription approaches that can generalize across arbitrary number of speakers, diverse settings (formal vs. informal conversations), meeting duration, wide-variety of acoustic scenarios and different recording configurations. 
Novelties with respect to C7DASR include: i) the addition of NOTSOFAR-1, an additional office/corporate meeting scenario, ii) a manually corrected Mixer 6 development set, iii) a new track in which we allow the use of large-language models (LLM)
iv) a jury award mechanism to encourage participants to explore also more practical and innovative solutions.
To lower the entry barrier for participants, we provide a standalone toolkit\footnote{\href{https://github.com/chimechallenge/chime-utils/}{github.com/chimechallenge/chime-utils/}} for downloading and preparing such datasets as well as performing text normalization and scoring their submissions.
Furthermore, this year we also provide two baseline systems, one directly inherited from C7DASR and based on ESPnet and another one developed on NeMo and based on NeMo team submission in last year C7DASR. 
Baseline system results suggest that the addition of the NOTSOFAR-1 scenario significantly increases the task's difficulty due to its high number of speakers and very short duration.
\end{abstract}

\section{Introduction}\label{sec:into}

The CH\-iME-8 DASR (C8DASR) challenge builds directly upon the previous C7DASR~\cite{cornell2023chime}, thus focusing on generalizable automatic speech recognition (ASR) and diarization with distant microphone devices. 
Such research direction is of high practical interest for a multitude of applications that go beyond mere meeting transcription. 
For example, the development of LLMs has led to significant advancements in chatbots, paving the way for new speech-enabled machine interaction possibilities. These include more useful speech-enabled assistants that can be better integrated into everyday tasks and professional environments via \eg\ meeting summarization and spoken information retrieval.
Moreover, transcribing ``speech-in-the-wild'' conversations in a robust way can be regarded as one of the key machine listening problems. In fact it requires to solve, at least implicitly, the notoriously difficult cocktail party problem and encompasses a suite of different speech processing tasks: voice activity detection (VAD) and diarization, speech separation, ASR and language modeling (LM).

Research towards this important direction has been fostered by several challenges and datasets that featured long-form conversations between several speakers. 
Such efforts can be traced back to the early 90s with the release of the Switchboard dataset, which featured telephone conversations, and then with AMI~\cite{carletta2005ami} and ICSI~\cite{janin2003icsi} corpora in the early 2000s, which instead consisted of office meeting scenarios recorded by far-field devices. Several others followed in the 2010s with Mixer 6~\cite{brandschain2010mixer}, Sheffield Wargames corpus~\cite{fox2013sheffield}, DiPCo~\cite{van2019dipco}, CH\-iME-5~\cite{barker2018fifth} and, more recently, in the past $4$ years with CH\-iME-6~\cite{watanabe2020chime}, Alimeeting~\cite{yu2022m2met} and Ego4D~\cite{grauman2022ego4d}. 



All these previous works focused on one particular scenario of interest, such as \eg\ office meetings (AMI, Alimeeting), dinner parties (CH\-iME-5/6, DiPCo) or ego-centric videos (Ego4D). On the other hand, in actual applications, it is desirable that a transcription system is robust across different domains and can handle a wide variety of conditions, such as the number of total speakers, meeting duration, and signal-to-noise ratio (SNR). On top of that, it is also desirable in many instances that such a system can be easily deployed across different recording setups and devices so that specialized ad-hoc devices are not required.  

The past C7DASR challenge was thus created to be an extension of the CH\-iME-6 challenge but with a special focus on such generalization problem. The main novelties with respect to CH\-iME-6 were: i) data was expanded to three different scenarios (CH\-iME-6, DiPCo~\cite{van2019dipco} and Mixer 6~\cite{brandschain2010mixer}), 
ii) expanded training material to include popular open source datasets (\eg LibriSpeech~\cite{panayotov2015librispeech}) and large-scale pre-trained models (\eg\ WavLM~\cite{chen2022wavlm}) iii) a novel ranking metric: diarization-attributed word error rate (DA-WER) to encourage participants to produce reasonable speaker-attribution and segmentation on top of recognition. 

C7DASR had a total of $9$ teams participating but succeeded only partially in some of the scientific goals that were set at the time of its organization. 
In particular, we saw all participants adopting end-to-end (E2E) automatic speech recognition (ASR) models. This is in stark contrast to the previous CH\-iME-6 challenge, where, instead, most of the teams relied on hybrid ASR systems. 
This shift was mainly due to the fact that we allowed the use of pre-trained models and, at the same time, relaxed the rules about allowed language modeling (LM) technique. 
Participants' submissions also demonstrated that it is possible to devise a single system that is array-topology agnostic and performs quite well across the scenarios considered. When compared with CH\-iME-6 challenge results, the best main track system~\cite{wang2023ustc} achieved around 30\% relative reduction in concatenated minimum permutation word error rate (cpWER). This was achieved by a remarkably effective diarization system~\cite{yang2023neural}, which improved a lot even over the highly effective target-speaker voice activity detection (TS-VAD)~\cite{medennikov2020target} approach. 

\vspace{-0.2cm}
\section{Motivation}\label{sec:motivation}

This said, C7DASR had, in our opinion, several shortcomings. Notably, most submissions (with \cite{boeddeker2023multi, karafiat2023but} being the only notable exceptions), were largely based on ensembling techniques, making many of the proposed approaches unpractical for real-world deployment. 
Another issue was that evaluation data was not really blind. Among the three scenarios only Mixer $6$ was partially blind, as its evaluation set data was manually re-annotated to feature both interviewer and subject transcriptions. This data however had been already available to the public for more than a decade. 
Moreover, the diversity in terms of the number of speakers across the three scenarios was still limited ($2$ to max $4$ speakers), and recent trends such as LLM integration~\cite{park2023enhancing} were not considered out of fear of raising the computational entry barrier too high for some teams. 

This year, the C8DASR challenge attempts to address these pitfalls by introducing several novelties.

\noindent
\textbf{More diverse scenarios.} We introduce a further scenario, NOTSOFAR-1~\cite{vinnikov2024notsofar}, which is shared with this year concurrent CH\-iME-8 NOTSOFAR-1 challenge~\cite{vinnikov2024notsofar}. 
NOTSOFAR-1 features office meetings between $4$ to $8$ participants of short duration ($\sim$ $6$\,mins) captured by a single far-field array device. 
As such, it adds significant diversity to the challenge scenarios, as can be seen in Table~\ref{tab:dset_overview}. On one hand, systems have to deal with long meetings with $4$ participants (CH\-iME-6), on the other, NOTSOFAR-1 features up to $8$ participants and a short duration, thus significantly complicating the diarization speaker counting component, as we will see in Sec.~\ref{sec:results}. 

An additional rationale for adding such a scenario is that it allows the direct comparison of C8DASR and CH\-iME-8 NOTSOFAR-1 challenge submissions and, thus, how generalist, array-agnostic systems (C8DASR) compare with domain and device-specialized ones. This is an interesting research question whose answer might not be obvious, as a generalist system may have the advantage of being able to leverage more training data compared to a specialized one and might actually generalize better in the case of significant train-test mismatches as it relies on fewer assumptions.   

\noindent
\textbf{Better data partitioning and annotation.}
Some datasets such as DiPCo and Mixer 6 do not offer, by default, long-form conversational data for training.
Such data is only available for development or evaluation.
Moreover, Mixer 6 had manually checked annotation only for evaluation, while the development set was automatically annotated using close-talk devices. 
In C8DASR, we manually re-annotated the development set of Mixer 6. 
Moreover, we split this latter and DiPCo development set in a separate training and development parts, so that we can have an official development set which can be used to assess generalization of submitted systems. 
In C7DASR, this was not possible as many participants were trained with the development data.

\noindent
\textbf{Better ranking metric.}
As said, C7DASR used DA-WER as the ranking metric which is a variant of cpWER where the permutation is determined by the optimal diarization mapping. 
However, in practice with a reasonable diarization system DA-WER and cpWER are basically equivalent, and, worse, again depending on the diarization system used, does not care really about segmentation but only about the overall speaker-id assignment. 
As such, this year, we adopt time-constrained cpWER (tcpWER)~\cite{von2023meeteval} as the ranking metric, which is a recently proposed metric that introduces temporal constraints in the Levenshtein distance computation and thus allows us to account for reasonable segmentation on top of speaker attribution and words recognition. 


\noindent
\textbf{Encouraging practical and innovative approaches.} To foster efforts also towards practical and innovative approaches, we introduce a special jury award, which is independent from the final ranking of the system. 
Since participants this year could submit up to 4 systems per track, they can explore both ``performance squeezing'' approaches as well as more pragmatic ones as both directions are important. 
This award is assigned by a pool of expert reviewers based on the submitted systems technical papers and systems metadata (in the form of a YAML file). 


\noindent
\textbf{Exploring how and if LLMs can be leveraged.}
Recent work suggests that LLMs can be leveraged effectively to improve meeting transcription, for example, regarding diarization~\cite{park2023enhancing} and ASR~\cite{yu2024connecting}. This year, we have a dedicated second track where participants can explore such interesting directions, together with more lightweight approaches for which the first track exists.

\noindent
\textbf{Lowering the entry bar for newcomers.} 
Joint ASR and diarization of long-form conversational speech with multiple devices is a very complicated task which is often addressed, as said, by a pipeline which consists of different components such as diarization, separation and ASR. 
As such, the entry bar for this challenge is quite high already. To make things worse, C7DASR and C8DASR feature different scenarios, and thus, lots of effort needs to be spent in the data preparation stage in order to parse the datasets and prepare manifest files for training \eg\ ASR or diarization components.  

This year, we spent a lot of effort trying to reduce participants' effort on this side as much as we could. 
For example, we built a new toolkit \texttt{chime-utils} that handles automatic downloading of CH\-iME-6, DiPCo, and NOTSOFAR-1 with a single line of code. 
Moreover, we convert all scenarios audio files and annotation to follow the same format and organization of CH\-iME-6 such that participants effort for data preparation and parsing can be minimized. 
Within \texttt{chime-utils} we also offer many data preparation tools for common speech processing toolkits: ESPnet~\cite{watanabe2018espnet}, Kaldi~\cite{povey2011kaldi} and K2 via Lhotse~\cite{zelasko2021lhotse}, as well as scoring scripts by interfacing with Meeteval~\cite{von2023meeteval}. 
Our hope is that this effort can also be beneficial to the community even after the end of this challenge as it could make datasets such as CH\-iME-6 and NOTSOFAR-1 more easy to experiment with in the future. 

Finally, we also offer two baselines systems, one implemented using ESPnet~\cite{watanabe2018espnet} and another one in NeMo~\cite{nemo_2019}. 
The ESPnet one is an updated version of C7DASR baseline, where the clustering threshold and diarization system hyper-parameters have been re-tuned to account for the newly added NOTSOFAR-1 scenario. 
The NeMo one is instead built on top of C7DASR NeMo team submission~\cite{park2023chime}. 
Both baselines are described in detail in Section~\ref{sec:baseline} and are built in a modular fashion, so that participants can re-use existing components and focus only on particular parts of the pipeline, \eg\ the diarization system or target speaker separation. 



\begin{table}[h]
\centering
\footnotesize
\caption{C8DASR scenarios diversity overview.}\label{tab:dset_overview}
\adjustbox{max width=\linewidth}{
\begin{tabular}{@{}llrrr@{}}
\toprule   
 \textbf{Scenario} & \textbf{Setting} & \textbf{Number of} & \textbf{Recording} & \textbf{Avg. }  \\
  & & \textbf{Speakers}\, & \textbf{Setup}\,\,\,\,\, & \textbf{Duration}\,\,\,\, \\ 

CH\-iME-6 & dinner party & 4 & 6 linear arrays & $\sim$2h-2h\,30\,mins \\ 
DiPCo & dinner party & 4 & 5 circular arrays & $\sim$20-30\,mins \\ 
Mixer 6 & 1-to-1 interview & 2 & 10 heterogeneous devices & $\sim$15\,mins \\ 
NOTSOFAR-1 & office meeting & 4-8 & 1 circular array & $\sim$6\,mins \\ 
\bottomrule
\end{tabular}
}
\end{table}

\setlist{
    topsep=0pt,
    partopsep=0pt,
    itemsep=0pt,
    parsep=0pt
}

\section{Datasets}

C8DASR features four core scenarios, all in English, with one additional scenario compared to the previous C7DASR challenge. 
Participants' systems are benchmarked on these four scenarios evaluation sets but, as in C7DASR, they can also use external datasets and pre-trained models from a predefined list available in the challenge website. 
As said, the use of pre-trained models such as WavLM was especially found to be key in the past challenge by all participants in order to obtain reasonable performance. 
These include meeting datasets such as AMI, VoxCeleb 1\&2~\cite{nagrani2017voxceleb} which is important for training speaker-id discriminative models but also clean speech (e.g. LibriSpeech) and noise-only datasets (Audioset~\cite{gemmeke2017audio}, SINS~\cite{dekkers2017sins}) in order to allow the creation of synthetic data. 
Regarding pre-trained models we include popular self-supervised training models (WavLM, wav2vec 2.0~\cite{baevski2020wav2vec}, HuBERT~\cite{hsu2021hubert}) as well as the recently released weakly supervised ASR model OWSM~\cite{peng2023reproducing}, other models such as ECAPA-TDNN~\cite{desplanques2020ecapa}, TitaNet~\cite{koluguri2022titanet} for speaker-id embedding extraction, Brohuaha~\cite{lavechin2023brouhaha} and MarbleNet~\cite{nguyen2022marblenet} for VAD to name a few. 
Contrary to C7DASR this year we also allow to use Whisper~\cite{radford2022robust}. In fact from ~\cite{cornell2023chime} results, it is likely that our evaluation data was not included in its training material since its performance is competitive with a WavLM-based ASR model. 
Participants could also propose to add additional pre-trained models and external datasets up to a certain date after the challenge beginning.

The $4$ core scenarios are described more in detail thereafter and their statistics are summarized in Table~\ref{tab:class_stats} while their characteristics are in Table~\ref{tab:dset_overview}.

\vspace{-1em}
\subsection{CH\-iME-6}
CH\-iME-6 consists of recordings of dinner parties between $4$ participants in a home environment across different rooms i.e. kitchen, living and dining rooms with participants free to roam across them. 
Far-field speech is captured via $6$ Kinect array devices with $4$ microphones each but, for annotation and training purposes also on-speakers binaural close-talk microphones are available. 
Compared to the other scenarios it features ``causal''-style informal conversations and high environmental noise as participants for example cook or dine together. 
Compared to C7DASR, we revert here to the original CH\-iME-6 \texttt{train}, \texttt{dev}, and \texttt{eval} splitting.   

\vspace{-1em}
\subsection{DiPCo}
DiPCo also features a dinner party scenario between 4 participants. 
However, compared to CH\-iME-6 all recordings take place in a single room, features arguably less informal conversations and, on average, have higher SNR. 
Recordings are captured by $5$ far-field devices each with a 7-mic circular array (6+1 microphone in the center) and by close-talk on person lapel microphones. These latter have far-less cross-talk and noise than CH\-iME-6 ones. 
DiPCo originally consists of $10$ sessions: $5$ \texttt{dev} and $5$ \texttt{eval}. 
As said in Section~\ref{sec:motivation}, we further split the original \texttt{dev} set into a \texttt{train} (3 sessions) and a \texttt{dev} partition ($2$ sessions) with approximately the same total duration. 

\vspace{-1em}
\subsection{Mixer 6}
Mixer 6 consists of $2$-speakers sessions (sampled at $16$\,kHz) recorded by $10$ different far-field recording devices and $3$ close talk devices. 
Each session consists of an interview part between an interviewer and a subject as well as a telephone call and a short prompt reading. 
In C8DASR, for \texttt{dev} and \texttt{eval} purposes, we make use only of the interview portion which is the only conversational speech portion in each session. 
Originally~\cite{brandschain2010mixer} annotation is only available for the subject. To obtain also annotation for the interviewer, in C7DASR, we used an automatic procedure and checked the annotation manually for the \texttt{eval} set only. 
For C8DASR we also performed manual check for the \texttt{dev} set. 
Originally Mixer 6 lacks fully transcribed conversational data for training, i.e., in C7DASR we provided for training two splits \texttt{train\_intv} and \texttt{train\_calls} where only the annotation for the subject is available.
For this reason, this year, we further split the \texttt{dev} set into a \texttt{train} and a \texttt{dev} set so that participants have some limited conversational data to fine-tune their systems. 
All the recordings take place into two rooms with one room used in all \texttt{train} and \texttt{dev} split and one unseen room in \texttt{eval}. 

\vspace{-1em}
\subsection{NOTSOFAR-1}

The original NOTSOFAR-1 features $315$ recordings of very short ($\sim 6$\,mins) office meetings between $32$ unique participants in 30 different meeting rooms. Meetings are between $4$ up to $8$ speakers and most are started by a professional actor whose task is to ``jumpstart'' and guide the conversation around a certain topic. 
Each meeting is captured by up to $7$ commercially available far-field array devices. These include $4$ tabletop circular devices with $7$ microphones each and $3$ linear array devices. For these latter however only monaural signals after in-device acoustic front-end processing is made available. 
In addition to far-field devices on-person close-talk headset devices with low cross talk are made available for \texttt{train} and \texttt{dev} splits. 
Ground truth annotation is available in the form of JSON files. Such annotation also includes word level alignment for each speaker utterance as well as metadata such as the meeting topic. Transcriptions were obtained with a multi-judge system as described in~\cite{vinnikov2024notsofar}. 
A more detailed description is available in~\cite{vinnikov2024notsofar}.

This data is used in C8DASR and CH\-iME-8 NOTSOFAR-1 tasks by considering just one device at a time: i.e. we consider as a unique session the recording of a meeting from each circular array. Thus, the number of sessions is significantly higher than the 315 total meetings (see Table~\ref{tab:class_stats}). This choice was made due to the fact that having just one far-field device is a highly practical occurring situation, and it is thus of great interest for many application scenarios. 
Also, as previously mentioned, we reformat and reorganize the data and annotation such that it has the same format as CH\-iME-6 and DiPCo to make it easier for participants the parsing of the $4$ different datasets. 
As seen in Table~\ref{tab:class_stats}, NOTSOFAR-1 data is organized into two training sets: \texttt{train} and \texttt{train\_sc}. This latter is single-channel only data from the aforementioned linear array devices and is thus included only for training purposes.



\vspace{-1em}
\begin{table}[h]
\centering
\footnotesize
\caption{CH\-iME-8 DASR core datasets statistics overview. 
We report the number of utterances, speakers, and sessions, as well as silence (sil), single-speaker speech (1-spk) and overlapped speech (ovl) ratios over the total duration. 
}\label{tab:class_stats}
\adjustbox{max width=\linewidth}{
\begin{tabular}{@{}llrrrrSSS@{}}
\toprule   
 \textbf{Scenario} & \textbf{Split} & \textbf{Size (h)} & \textbf{Utts} & \textbf{Spk.} & \textbf{Sess.} & \textbf{sil (\%)} & \textbf{1-spk (\%)} & \textbf{ovl (\%)}   \\
\midrule
\multirow{3}{*}{\textbf{CH\-iME-6}} & \multirow{1}{*}{train} &  \multirow{1}{*}{40:05} &  79967   & \multirow{1}{*}{32} & \multirow{1}{*}{16} & 22.6  & 52.7  & 24.7  \\
& \multirow{1}{*}{dev} &  \multirow{1}{*}{4:27} &  7437  & \multirow{1}{*}{8} & \multirow{1}{*}{2} & 13.1  & 43.4  & 43.5  \\
& \multirow{1}{*}{eval} &  \multirow{1}{*}{5:12} &  11028  & \multirow{1}{*}{8} & \multirow{1}{*}{2} & 21.3  &  52.0  &  26.7  \\
\midrule

\multirow{3}{*}{\textbf{DiPCo}} & \multirow{1}{*}{train} &  \multirow{1}{*}{1:12} & 1379  & \multirow{1}{*}{8} & \multirow{1}{*}{3} & 8.3 & 72.0 &  19.6 \\
& \multirow{1}{*}{dev} &  \multirow{1}{*}{1:31} &  2294  & \multirow{1}{*}{8} & \multirow{1}{*}{2} &  7.4 & 61.9  & 30.6 \\
& \multirow{1}{*}{eval} &  \multirow{1}{*}{2:36} & 3405  & \multirow{1}{*}{16} & \multirow{1}{*}{5} & 9.4  &  65.7 & 24.9  \\

\midrule
\multirow{4}{*}{\textbf{Mixer 6}}  & train calls &  36:09 & 27280  & 81 & 243 & {--}   & {--}  &    {--}  \\
 & train intv & 26:57  &  29893  & 77 & 189 & {--} &  {--} &  {--}    \\
  & train & 6:13  &   3785 & 19 & 24 & 8.6 & 73.3  &  18.0    \\
& dev &  8:56 & 5903 & 22 & 35 & 8.4 & 72.1 &  19.5   \\
& eval & 5:45 & 5115  & 18 & 23 &  2.4 & 83.6 & 13.9 \\
\midrule
\multirow{4}{*}{\textbf{NOTSOFAR-1}} & \multirow{1}{*}{train} &  \multirow{1}{*}{14:43} & 101301  & \multirow{1}{*}{14} & \multirow{1}{*}{379} & 6.0 & 62.3 &  31.7  \\

& \multirow{1}{*}{train\_sc} &  \multirow{1}{*}{53:43} & 139913  & \multirow{1}{*}{14} & \multirow{1}{*}{526} & 5.9  & 62.4  & 31.7 \\

& \multirow{1}{*}{dev} &  \multirow{1}{*}{13:25} & 24238  & \multirow{1}{*}{11} & \multirow{1}{*}{130} & 15.6  &  67.7  & 16.7 \\
& \multirow{1}{*}{eval} &  \multirow{1}{*}{16:29} & 38662  & \multirow{1}{*}{12} & \multirow{1}{*}{160} & 5.6  & 64.7  & 29.6  \\
\bottomrule
\end{tabular}
}
\vspace{-0.4cm}
\end{table}

\vspace{-0.2cm}
\section{Challenge Tracks \& Rules}
C8DASR challenges participants to produce transcriptions of long-form recordings recorded by one or more far-field recording devices.
The transcriptions need to have speaker-attribution and segmentation at the utterance level and should be in the form of a JSON SEGment-wise Long-form Speech Transcription annotation~\cite{von2023meeteval} (segLST) file\footnote{described in \href{https://www.chimechallenge.org/current/task1/submission}{chimechallenge.org/current/task1/submission}} which is the same format as used in the past C7DASR challenge.

\vspace{-1em}
\subsection{Evaluation Tracks}


This year, we don't have any oracle diarization track, contrary to the previous CH\-iME-5/6 and C7DASR challenges. 
This is mainly for two reasons: i) it makes the challenge more prone to cheating, defeating the purpose of having NOTSOFAR-1 scenario blind, thus possibly discouraging participation ii) in C7DASR challenge we found little additional insight from the oracle diarization sub-track.
Instead, there are two identical tracks in which participants have to perform diarization+ASR, which only differ by the allowed external pre-trained models:

\begin{itemize}
    \item \textbf{Constrained LM track.} This is equivalent to last year C7DASR main track. 
    \item \textbf{Unconstrained LM track.} Equivalent to the one above, but participants can also leverage additional LLMs pre-trained models including Llama2~\cite{touvron2023llama}, Megatron~\cite{shoeybi2019megatron} etc.\footnote{specified in \href{https://www.chimechallenge.org/current/task1/rules\#external\_lms}{chimechallenge.org/current/task1/rules\#external\_lms}}. 
\end{itemize}

\noindent
In both tracks participants can submit up to $4$ systems and are ranked according to tcpWER macro-averaged across all scenarios, computed with a collar of $5$ seconds. 
For evaluation, due to the fact that the $4$ core scenarios have small discrepancies in the annotation,
we employ a text normalization strategy borrowed from Whisper with some modifications and fixes.  
In detail, we remove Whisper number normalization and keep instead numbers as words (i.e. 20\$ will be converted to ``twenty dollars'') while common non-verbal speech sounds e.g. ``uhm'', ``uhhh'', ``ah'' as well as laughs are removed. All text is converted to lowercase and abbreviations such as ``mr'', ``prof'', ``goin'' are expanded. 
During data preparation via \texttt{chime-utils} we make these text normalizations, which will be used for scoring explicitly to the participants and generate both \texttt{transcription} folder and a \texttt{transcription\_scoring} one containing respectively the original annotation and the text normalized one. 
Participants are free to use whatever normalization strategy they prefer during training. 

\vspace{-1em}
\subsection{Rules}
Rules are largely the same as the previous C7DASR challenge. The main rationale is to discourage automatic or manual domain identification or any use of a-priori information from the scenarios. This is to prevent the systems from being ``by design'' limited in their generalization capability. The only crucial difference is that this year, since we provide official training and development partitions for DiPCo and Mixer 6, we don't allow training on the development set. 
Instead, in C7DASR, participants were allowed to re-arrange such development data with training data. This however complicated analysis on how well submitted systems were able to generalize to the evaluation set. Another difference is that we allow the use of some pre-trained LLMs but only in the Unconstrained LM track. Full description of the task rules is available in the challenge website\footnote{see \href{https://www.chimechallenge.org/current/task1/rules}{chimechallenge.org/current/task1/rules}}

\begin{table*}[!h]
    \caption{CH\-iME-8 DASR ESPnet and NeMo baselines diarization results in terms of DER (\%) and its components missed speech (MS), false-alarm speech (FA) and speaker confusion (SC). We also report speaker counting errors (\#SPK) in terms of percentage (\%) over total speakers of missed speakers (MS) and false alarm speakers (FA).  
    We highlight best figures between the two baselines for each scenario.
   }
    \vspace{-0.2cm}
    \label{tab:diarization_results}
    \centering
  
    \setlength{\tabcolsep}{7pt}
    \begin{tabular}{clcccc|cc|cccc|cc}
    \hline
          &  & \multicolumn{6}{c}{\textbf{Dev}} &   \multicolumn{6}{c}{\textbf{Eval}} \\
         \textbf{Baseline}  &  & \multicolumn{4}{c}{\textbf{diarization (\%)}} & \multicolumn{2}{c}{\textbf{\#SPK (\%)}} &  \multicolumn{4}{c}{\textbf{diarization (\%)}} &  \multicolumn{2}{c}{\textbf{\#SPK (\%)}} \\
        \cmidrule{3-14}
        \textbf{System} &  \textbf{Scenario} & MS & FA  &  SC  & DER & MS & FA & MS & FA  &  SC  & DER & MS & FA   \\ 
        \toprule
         \multirow{4}{*}{ESPnet} & CH\-iME-6 & \textbf{8.0} & 19.2 & 36.1 & 63.3 & 0.0 & 50.0 & \textbf{11.0} & \textbf{12.5} & 36.3 & 60.0   & 50.0 & \textbf{0.0} \\
                                &  DiPCo &  \textbf{10.8}  & 15.5 &  40.3  & 66.6 & 37.5 & \textbf{0.0}   & \textbf{6.4} & \textbf{5.5}  &  \textbf{8.6} & \textbf{20.5} & 10.00 & \textbf{30.0} \\
                                & Mixer 6 &  \textbf{3.0}  & 7.9 & 4.6  & \textbf{15.5} &  \textbf{0.0} & 32.8 & \textbf{5.8}  & \textbf{0.9} & 3.6 & \textbf{10.3} & \textbf{0.0} & 32.5 \\ 
                                 & NOTSOFAR-1 &  \textbf{8.8}  &  \textbf{7.4} & \textbf{14.4}  & \textbf{30.6} & \textbf{19.6} & 0.8 & \textbf{3.3} & \textbf{1.1}  & \textbf{8.5} & \textbf{12.8} & \textbf{15.4} & \textbf{0.5} \\ 
                             
    \midrule
    \multirow{4}{*}{NeMo} & CH\-iME-6 &  21.3  & 11.9  & 9.8  & \textbf{43.0} & \textbf{0.0} & \textbf{25.0} & 18.3 & 15.8 & \textbf{22.6} & \textbf{56.7} & \textbf{0.0} & 37.5 \\
                                &  DiPCo & 16.4  & \textbf{11.3} & \textbf{19.6} & \textbf{47.3} & \textbf{0.0} &  62.50 & 10.1   & 10.8 & 15.3 & 36.2 & \textbf{0.0} & \textbf{30.0} \\
                                & Mixer 6 &  10.6  & \textbf{3.7} & \textbf{2.1}  & 16.5 & \textbf{0.0} & \textbf{0.0} &  8.4 & 4.31 & \textbf{0.7} & 13.4 & \textbf{0.0}  & \textbf{0.0}  \\ 
                                & NOTSOFAR-1 &  9.76  &  8.6 & 14.0  & 32.4 & 32.9 & \textbf{0.0} &   10.4 & 11.7 & 24.8 & 47.00 & 40.8 & 0.9 \\

    \bottomrule
    \end{tabular}
    \vspace{-1em}
\end{table*}


\begin{table}[!htbp]
    \caption{Top panel: CH\-iME-8 DASR ESPnet and NeMo baselines overall results in terms of cpWER (\%) and tcpWER (\%). We highlight best figures between the two baselines for each scenario. Bottom panel: figures obtained by last year ESPNet C7DASR baseline.}
    \vspace{-0.2cm}
    \label{tab:final_results}
    \centering
    \setlength{\tabcolsep}{7pt}
    \begin{tabular}{clcccc}
    \hline
        &  & \multicolumn{2}{c}{\textbf{Dev}} &   \multicolumn{2}{c}{\textbf{Eval}} \\
        \textbf{Baseline }   & &  \multicolumn{2}{c}{WER (\%)} & \multicolumn{2}{c}{WER (\%)} \\ 
         \cmidrule{3-6}
        \textbf{System} & \textbf{Scenario} & cp & tcp &  cp  & tcp \\ 
      
        \toprule
         \multirow{4}{*}{ESPnet} & CH\-iME-6 & 79.2 & 88.6 & 91.8 &  99.1 \\
                                & DiPCo &  90.9  &  98.3 & 52.8  &  \textbf{56.6} \\
                                & Mixer 6 & \textbf{23.4}  & \textbf{23.9} & 42.0  &  43.8 \\ 
                                 & NOTSOFAR-1 &  \textbf{42.4}  & \textbf{46.2} & \textbf{48.5}  & \textbf{50.7} \\ 
                                 \cmidrule{2-6}
                                & Macro & 59.0 & 64.2 & 58.8  & 62.6   \\ 
    \cmidrule{1-6}
    \multirow{4}{*}{NeMo} & CH\-iME-6 &  \textbf{52.2}  &  \textbf{56.5}  & \textbf{67.7} & \textbf{73.8} \\
                                & DiPCo & \textbf{72.3} & \textbf{75.8}  &  \textbf{54.6} & 57.1 \\
                                & Mixer 6 &  17.9   &  19.4  & \textbf{22.3} & \textbf{23.1}  \\ 
                                & NOTSOFAR-1 &  55.5  & 61.0 &  67.2 & 72.0 \\ 
                                 \cmidrule{2-6}
                                & Macro & \textbf{49.6}  & \textbf{53.2}  & \textbf{52.9}  &  \textbf{56.5} \\   
    \midrule
     \midrule
       \multirow{3}{*}{C7DASR} & CH\-iME-6 & 60.8  & 65.7   & 73.7 &  85.2 \\
                                & DiPCo &   38.0 & 38.9  & 52.4  &  58.4 \\
                                & Mixer 6 & 20.7  & 21.5  & 31.7  &  32.2  \\ 
         
    \bottomrule 
    \end{tabular}
    \vspace{-2em}
\end{table}

\section{Baseline Systems}\label{sec:baseline}

\begin{figure}
    \centering


  
\begin{tikzpicture}[every node/.style={font=\footnotesize}]
\tikzset{>=latex}
\tikzstyle{branch}=[{circle,inner sep=0pt,minimum size=0.3em,fill=black}]
\tikzstyle{block}=[
    draw,
    text depth=0pt,
    thick, 
    rectangle,
    text centered,
    minimum height=3.2ex,
    rounded corners=0.3em,
    fill=black!6,
    inner sep=0.7em,
    ]
\tikzstyle{arrow}=[{}-{>}, thick]
\tikzset{
  wav/.pic={
    \begin{scope}[x=0.2em,y=0.07em]
      \coordinate (-west) at (-1,0);
      \coordinate (-east) at (18,0);
      \coordinate (-south) at (8.5,-8);
      \coordinate (-north) at (8.5,8);
      \foreach \x/\y in {0/1.2,1/2,2/4,3/7,4/4,5/2,6/1.2,7/2,8/1.2,9/2,10/4,11/7,12/4,13/7,14/1.2,15/2,16/5,17/1.2} {
          \draw[line width=0.12em] (\x,\y) -- (\x,-\y);
      }
    \end{scope}
  }
}

\definecolor{tss}{HTML}{9dc3e5}
\definecolor{mcd}{HTML}{ffecb2}
\definecolor{asr}{HTML}{a9d18e}

\node[block,align=center,fill=tss] (tss) at (0, 0) {Target Speaker\\Separation};
\node[block,align=center,anchor=north,fill=mcd] (mcd) at ($(tss.south)+(0,-1.8em)$) {Multi-Channel\\Diarization};

\pic (wavOne) at ($(tss.west) + (-6.1em,-1em)$) {wav};
\pic (wavTwo) at ($(tss.west) + (-7.1em,0)$) {wav};
\pic (wavThr) at ($(tss.west) + (-8.1em,1em)$) {wav};

\coordinate (tmp) at ($(tss.south) + (-0.7em,0)$);
\draw[arrow] (mcd.north-|tmp) -- node[right]{Diarisation} (tmp);
\draw[arrow] (wavOne-east) -- (tss.west|-wavOne-east);
\draw[arrow] (wavTwo-east) -- (tss.west|-wavTwo-east);
\draw[arrow] (wavThr-east) -- (tss.west|-wavThr-east);

\draw[arrow] ($(tss.west|-wavOne-east)$) +(-1.9em,0) node[branch] {} |- ++($(mcd)-(tss)$);
\draw[arrow] ($(tss.west|-wavTwo-east)$) +(-1.5em,0) node[branch] {} |- ++($(mcd)-(tss)$);
\draw[arrow] ($(tss.west|-wavThr-east)$) +(-1.1em,0) node[branch] {} |- ++($(mcd)-(tss)$);

\node[anchor=east] () at ($(wavOne-west) + (0,0)$) {$x_1(t)$};
\node[anchor=east] () at ($(wavTwo-west) + (0,0)$) {$x_2(t)$};
\node[anchor=east] () at ($(wavThr-west) + (0,0)$) {$x_m(t)$};

\pic (wavOut) at ($(tss.east) + (2em,0em)$) {wav};
\node [above] () at (wavOut-north) {$x_{\mathrm{t}}(t)$};

\node[block,align=center,anchor=north,fill=asr] (asr) at ($(wavOut-south)+(0,-1.2em)$) {ASR};

\draw[arrow] (wavOut-south) -- (asr);
\draw[arrow] (tss) -- (wavOut-west);
\draw[arrow] (asr.south) -- node[below=0.2em]{Transcriptions} +(0,-1em);

\end{tikzpicture}
  
    \caption{ESPNet and NeMo baseline systems basic overview.}
    \label{fig:baseline_scheme}
    \vspace{-2em}
\end{figure}
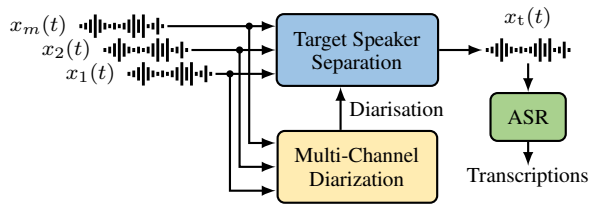

As said, this year we feature two baseline systems, one implemented in ESPnet and another in NeMo. 
All baseline systems follow the scheme outlined in Figure~\ref{fig:baseline_scheme}: first multi-channel diarization is performed, then the diarization output is used to perform guided source separation (GSS)~\cite{boeddeker2018front} and finally a monaural ASR is used to recognize each separated utterance. 


\vspace{-1em}
\subsection{ESPnet Baseline}
It is the same baseline as the one employed in C7DASR. 

\noindent
\textbf{Diarization system.} The multi-channel diarization system is a multi-channel extension of the $2.1$ Pyannote diarization pipeline~\cite{bredin2020pyannote}. It employs a local end-to-end neural diarization (EEND) module~\cite{bredin2021end} with a context size of 5\,s to exclude non speech regions and overlapped speech regions. Following, ECAPA-TDNN speaker-id discriminative embeddings are extracted on single-speaker regions, clustered and finally speech regions (including overlapped speech ones) are reassigned according to~\cite{bullock2020overlap}. 
Since many scenarios feature multiple devices, the local EEND module is also used here to perform microphone channel selection by selecting the channel among all available ones which has the most speaker activity. 

\noindent
\textbf{Target speaker separation.} 
We employ envelope variance (EV) channel selection~\cite{WOLF_EV_2014} followed by GSS. For this latter we use a GPU-accelerated implementation of GSS~\cite{Raj2022GPUacceleratedGS} that uses MIMO-WPE dereverberation~\cite{yoshioka2012generalization} followed by a GSS-driven MVDR~\cite{capon_mvdr} beamformer with a-posteriori maximum SNR channel selection~\cite{erdogan2016improved}.
The use of EV channel selection has been demonstrated to both speed up and improve results when combined with GSS~\cite{cornell2023chime}.

\noindent
\textbf{Automatic speech recognition.}
The ASR model is based on~\cite{chang2022end, masuyama2022end} and consists of a hybrid CTC/Attention transformer encoder-decoder model with WavLM-based features. 
It is trained on the full CH\-iME-6 \texttt{train} and Mixer 6 \texttt{train\_intv} and \texttt{train\_calls} splits, employing all microphones, including close-talk ones and GSS-enhanced data. Close-talk microphone data is also augmented following the CH\-iME-6 baseline augmentation scripts~\cite{watanabe2020chime}. 
\vspace{-1em}
\subsection{NeMo Baseline}
It is based on NeMo team C7DASR submission. The main difference compared to the ESPnet one is the diarization system which is based on the TS-VAD framework. 

\noindent
\textbf{Diarization system.}
The diarization pipeline consists of MIMO-WPE applied on 40\,s windows with 2\,s overlap, followed by channel clustering, which is performed using NME-SC clustering~\cite{park2019auto} of the spatial coherence matrix of all channels, computed across the whole meeting.
Afterwards, a multi-channel VAD based on MarbleNet is employed. 
Such VAD model is applied on each output channel from the channel clustering step and logits are fused by taking the max over all channels. 

Diarization relies on a novel TS-VAD attention-based multi-scale diarization decoder (MSDD)~\cite{park2022multi} model which employs multi-scale (3\,s, 1.5\,s, and 0.5\,s) TitaNet-large embeddings extracted on speech regions. The MSDD model has four-layer transformer architecture with a hidden size of $384$. 
Target speaker embeddings are estimated in inference using NME-SC clustering using the multi-scale TitaNet embeddings. MSDD is applied on each channel after the channel clustering step and results are aggregated via majority voting. 
Both VAD and MSDD models are trained on CHiME-6 training subset and simulated data using NeMo multi-speaker data simulator~\cite{park2023property} on Voxceleb1\&2 datasets~\cite{nagrani2017voxceleb}.

\noindent
\textbf{Target speaker separation.}
Same as in the ESPnet baseline but re-implemented in NeMo using Pytorch. 

\noindent
\textbf{Automatic speech recognition.}
The ASR model is based on NeMo Conformer-based transducer~\cite{gulati2020conformer}, which is fine-tuned using CH\-iME-6 and Mixer 6 training data after GSS pre-processing. 
During beam-search an n-gram LM is used. It is implemented via KenLM~\cite{heafield2011kenlm} and based on byte-pair-encoding (BPE) tokens obtained via SentencePiece~\cite{kudo2018sentencepiece}. This LM is trained on text data from CH\-iME-6 and Mixer 6 training splits.

\section{Baseline Results \& Discussion}\label{sec:results}

In Table~\ref{tab:final_results} top panel, we report each baseline results on the four core scenarios \texttt{dev} and \texttt{eval} sets in terms of tcpWER (which is the ranking metric) and cpWER. We also report the macro-average across all four core scenarios. 

First, it is interesting to compare results obtained by this year ESPnet baseline here and the C7DASR one (Table~\ref{tab:final_results}, bottom panel).
We can see that there is a significant degradation for CH\-iME-6 and Mixer 6 scenarios. 
Such performance reduction is mostly due to wrong total speaker counting as we will see later.
We can also observe that cpWER figures on e.g. Mixer 6 and DiPCo \texttt{eval} set and CHiME-6 \texttt{dev} set are marginally lower (i.e. on average by $\sim 2\%$) compared to last year ones. This is mostly because of the new text normalization strategy adopted now. In fact, as said, DA-WER and cpWER are practically equivalent as long as the diarization system can correctly attribute most of the speakers (which is the case for C7DASR baseline). 

The NeMo baseline fares a bit better overall due to the better diarization component and better ASR model. 
Especially its results are more consistent across Mixer 6 and the performance is significantly better in the CH\-iME-6 scenario. 
However, it is significantly worse on NOTSOFAR-1, possibly due to the fact that this pipeline, as said, is derived from NeMo C7DASR submission and, thus, it was designed to tackle mainly CH\-iME-6, DiPCo and Mixer 6. 
For example, as explained, the number of speakers is estimated via clustering TitaNet embeddings without any overlapped speech exclusion. This design choice may lead to degraded speaker counting performance is some situations like NOTSOFAR-1 where overlapped speech is significant, the speakers are numerous and the meeting very short.

In Table~\ref{tab:diarization_results} we report diarization and speaker counting results for the two baselines across the four core scenarios \texttt{dev} and \texttt{eval} sets. 
The reason for performance difference between the two baselines is evident when looking at diarization error rate (DER) speaker confusion (SC) and speaker counting missed speakers (\#SPK-MS) and false alarm speakers (\#SPK-FA). 
In fact, we can observe that the ESPnet baseline has more speaker counting errors relative to the NeMo one in the CH\-iME-6 scenario. The opposite is true instead for NOTSOFAR-1. 
It is also evident that by looking only at the DER value it is not always possible to assess which one of the two baselines is better on which scenario. 
For example, for CH\-iME-6 \texttt{eval} set the total DER value is quite close (60.0\% vs 56.7\%), however the ESPnet baseline has significantly higher SC and overall worse speaker counting, leading to much worse cpWER and tcpWER compared to the NeMo system. 

\section{Conclusions}
In this paper we presented the CH\-iME-8 DASR challenge task, which extends the previous edition C7DASR task by adding an additional scenario, a new track where participants can leverage LLMs and a jury award mechanism to encourage participants to focus also on practical and innovative approaches. This task focuses on generalizable multi-channel far-field meeting transcription and thus participants are challenged to develop ASR+diarization systems that can generalize across 4 different scenarios with very diverse characteristics. 
Results from the two baseline systems, one implemented in ESPnet and one in NeMo, suggest that one of the most challenging aspects is accurate total meeting speaker counting, as this component is the one responsible for most downstream recognition errors.

\section{Acknowledgments}
We thank the Linguistic Data Consortium (LDC) for providing free access to participants for the Mixer 6 Speech corpus during the whole duration of the challenge. We also want to thank NOTSOFAR-1 authors and in particular Alon Vinnikov, Amir Ivri and Shai Pe'er for the helpful discussions, collaboration and coordination between this year CH\-iME-8 DASR and NOTSOFAR-1 tasks.
Finally we thank also Jee-weon Jung for his helpful feedback for the debugging of \texttt{chime-utils} data generation.

\bibliographystyle{IEEEtran}
\footnotesize
\bibliography{refs}

\end{document}